\documentclass[12pt]{iopart}

\usepackage{iopams}  
\usepackage{cite}
\usepackage{graphicx}
\graphicspath{ {./figures/} }
\usepackage{epsfig}
\usepackage{epstopdf}
\usepackage{textcomp}
\usepackage{multirow}
\usepackage{fullpage}
\usepackage{float}
\usepackage{caption}
\usepackage{siunitx}
\usepackage{xcolor}
\usepackage{lmodern}
\usepackage[utf8]{inputenc}

\begin{document}

\title{Predictive modeling of solidification during laser additive manufacturing of nickel superalloys: Recent developments, future directions}

\author{Supriyo Ghosh\footnote{Guest Researcher.}}

\address{Materials Science and Engineering Division, National Institute of Standards and Technology, Gaithersburg, MD 20899, USA}
\ead{supriyo.ghosh@nist.gov}
\vspace{10pt}

\begin{abstract}
Additive manufacturing (AM) processes produce parts with improved physical, chemical, and mechanical properties compared to conventional manufacturing processes. In AM processes, intricate part geometries are produced from multicomponent alloy powder, in a layer-by-layer fashion with multipass laser melting, solidification, and solid-state phase transformations, in a shorter manufacturing time, with minimal surface finishing, and at a reasonable cost. However, there is an increasing need for post-processing of the manufactured parts via, for example, stress relieving heat treatment and hot isostatic pressing to achieve homogeneous microstructure and properties at all times. Solidification in an AM process controls the size, shape, and distribution of the grains, the growth morphology, the elemental segregation and precipitation, the subsequent solid-state phase changes, and ultimately the material properties. The critical issues in this process are linked with multiphysics (such as fluid flow and diffusion of heat and mass) and multiscale (lengths, times and temperature ranges) challenges that arise due to localized rapid heating and cooling during AM processing. The alloy chemistry-process-microstructure-property-performance correlation in this process will be increasingly better understood through multiscale modeling and simulation.
\end{abstract}

\section{Introduction}
The production of metallic parts via additive manufacturing processes (for recent reviews, see~\cite{Murr2012}, \cite{Frazier2014}, and \cite{Herzog2016}) such as laser powder bed fusion and direct metal laser sintering is growing rapidly to achieve the ever-increasing demand for improved strength and resistance to creep and fatigue for aerospace, defense, and medical applications~\cite{Reed2008,Attallah2016,Singh2017}. Ni$-$based superalloys in this context possess excellent mechanical properties at elevated temperatures, making them essential to the above sectors. However, there is a lack of confidence in the quality of additively manufactured parts due to the inherent multiscale and multiphysics problems associated with processing. On a macroscopic scale, i.e. millimeters, as the laser rasters across the powder bed, local regions begin to melt resulting in a molten pool of certain dimensions. The molten pool then undergoes a solidification, remelting and subsequent additional solidification processes driven by the resultant complex thermal history due to repeated passes of the laser. Temperature is highest at the top surface and varies along the depth, width and length of the melt pool. Therefore, the melt pool represents different local solidification conditions. As these local conditions vary, the microstructures within the solidified puddle also vary in different locations. The uncertainties in the quality of the parts arise precisely due to those location specific complex microstructures and due to unpredictable microstructure evolution paths, leading to beneficial or detrimental phases and microstructural anisotropies due to segregation and orientations~\cite{Suresh2016,Attallah2016,Murr2012,Francois2017,Ranadip2017}. These issues have received little attention, but they are important to consider during AM solidification.

The new classes of AM materials are substantially different than those produced by traditional casting, mechanical working and final machining mechanisms. This is due to the rapid solidification processing (RSP) during AM, leading to nonequilibrium segregation of the alloy elements in the molten pool, making the quantification of the resulting microstructures difficult. The prediction of material properties therefore becomes a serious hurdle, resulting in a lack of confidence in the quality of the final part. Some \SI{47}{\%} of manufacturers surveyed indicated that the uncertain quality of the final product was a barrier to adoption of additive manufacturing~\cite{Khairallah2016,King2014,King2015}. A predictive multiscale modeling framework could optimize the AM process parameters to improve the likelihood of producing qualified parts.

\begin{figure*}[ht]
\begin{center}
\includegraphics[scale=0.5]{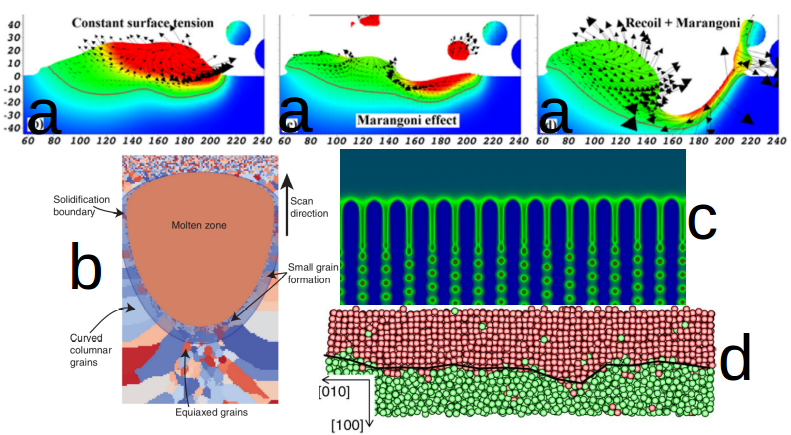}
\caption{(a) Typical melt pool shapes are shown for the consideration of different physics within the melt pool (reproduced from \cite{Khairallah2016}, with permission from Elsevier). (b) Typical solidification boundary is illustrated from which the local solidification conditions are estimated (reproduced from \cite{Rodgers2017}, open access). Melt pool solidify into columnar (on the side) and equiaxed (bottom) dendritic microstructures. (c) Phase-field simulation predicts a typical columnar dendritic morphology. The colors represent spatial solute concentration variation. Solute enriched droplets pinch off from the dendrite roots, where the secondary solid phases form. (d) Typical snapshot of a solid-liquid interface in thermal equilibrium during MD simulation is shown (reproduced from~\cite{Hoyt2003}, with permission from Elsevier).}\label{figure}
\end{center}
\end{figure*}

Due to the high temperature and small volume of the molten pool, \emph{in situ} measurements of the solidification conditions are difficult. Numerical simulations of the laser deposition process is a viable alternative to obtain the local solidification conditions in the melt pool. A finite element analysis (FEA) method, which includes heat transfer, fluid flow, Marangoni convection and other hydrodynamic effects~\cite{amberg2008,King2015}, can simulate realistic melt pool shapes as well as temperatures. Energy balance equations on the macroscale are solved in ABAQUS~\cite{Abaqus}/ANSYS~\cite{Ansys}/COMSOL~\cite{Comsol} environment to obtain the temperatures and geometries in the melt pool (refer to Fig.~\ref{figure}a). Laser processing parameters such as the laser power, speed, and size control the dimensions of the melt pool. These dimensions are important as they determine the density of the final parts through the solidification microstructures as well as through the cooling rates; for example, with increasing laser power or reduced laser speed, the melt pool becomes deeper~\cite{Khairallah2016,King2015}. Solidification conditions therefore change along the melt pool solid-liquid boundary and influence the resulting microstructure formation.

For the microstructure simulations, phase-field models~\cite{boettinger2002,chen2002,steinbach2009} have been the most popular choice. In these models, a scalar-valued order parameter field $\phi$ is introduced to distinguish the phases present in a microstructure: if liquid is defined by $\phi = -1$ and solid by $\phi = 1$, then the interface can be taken as the $\phi = 0$ contour. Therefore, explicit tracking of the interface is no longer needed, and one can simulate the complex solid-liquid interface in an efficient way. Phase-field equations of motion are essentially time-dependent partial differential equations, which are solved by efficient numerical methods to obtain the steady state composition, temperature and order parameter maps through the solution of diffusion of heat, composition, and order parameter fields. The microstructures simulated in this way correspond only to a particular location in the melt pool. During the course of phase-field simulations, melt pool often solidifies into columnar and/or equiaxed dendritic microstructures. These microstructures often contain defects such as microporosity and solidification and liquation cracks, microstructural anisotropies due to alloy chemistry, segregation and orientations of the solidifying dendrites, and residual stresses due to shrinkage during terminal solidification. These complex features within the microstructure collectively influence the properties and performance of the solidified material. These effects have yet to be explored for AM solidification. Simulation of these aspects may require considerable computation time. Fortunately, phase-field models are found to scale well in parallel environments (MPICH~\cite{Mpich}, OpenMP~\cite{Openmp} and CUDA~\cite{Cuda}) to run the simulation codes faster. The phase-field model parameters are approximated to behave quantitatively only at small-valued solidification conditions~\cite{Echebarria2004}. Better models are therefore needed to treat larger values of the melt pool solidification conditions, appropriate for AM.

For more accurate modeling of the melt pool solidification, realistic solid-liquid interface properties are also needed. Molecular dynamics (MD) simulation can provide the information regarding the interfacial energy anisotropy and the interfacial kinetic coefficient for metallic alloys in AM solidification regime. Such an efficient combination of finite element, phase-field, and molecular dynamics approaches could potentially capture the predictive AM microstructural evolution. The present overview is aimed at the following outstanding issues that arise due to the multiscale, multiphysics nature of the solidification in the melt pool.

\section{Macroscale: Finite element analysis}
FEA simulations determine the actual solidification conditions in the melt pool. Using a suitable commercial software, a FEA model generate the global temperature history during laser irradiation on a single/multi layer(s) of powder of finite thickness deposited on a solid substrate of the same base alloy~\cite{supriyo2017,Trevor2017}. Ni$-$based superalloys have been commonly used for the powder as well as the substrate properties in these simulations. Bulk material properties, such as latent heat, density, and specific heat, were estimated from CALPHAD-based thermodynamic calculations~\cite{Smith2016_1,Smith2016_2,Andersson2002}. To reduce computational time, the FEA mesh elements that interact with the laser beam were finely meshed, and a coarse mesh was used in the far-field. Heat input from the laser was approximated in the literature by a moving point, line or plane heat source~\cite{Rosenthal1946}, a double ellipsoidal volumetric heat source~\cite{Goldak1984}, or a Gaussian source~\cite{Guo2000,Suresh2016}. Both convective and radiative heat losses were considered in these studies. The temperature distribution as a function of time was obtained by solving equations for the conservation of energy. The computed results are generally visualized using the appropriate visualization module of the FEA software or using a custom MATLAB~\cite{Matlab}/PARAVIEW~\cite{Paraview} program. The temperature gradient at each FEA element was estimated by the magnitude along the Cartesian directions using the temperature values from the neighboring elements. The trailing edge of the melt pool in experiments represents the solid-liquid boundary (refer to Fig.~\ref{figure}b), which was approximated by the melting temperature isotherm. Solidification begins at this boundary and the resulting columnar microstructures grow roughly perpendicular to this boundary~\cite{supriyo2017,Trevor2017}. The curvature of this boundary introduces an angle, which correlates the beam speed with the local solidification velocity. The solidification boundary represents different temperature gradients and solidification rates. Typically in simulations, the temperature gradient varies between $\approx$ $10^5$ K m$^{-1}$ and $10^7$ K m$^{-1}$ and the solidification rate varies between $\approx$ $0.01$ m s$^{-1}$ and $0.5$ m s$^{-1}$ for a laser scan speed on the order of $1$ m s$^{-1}$~\cite{supriyo2017,Trevor2017}. Temperature gradient times the solidification velocity is the cooling rate. These local parameters were provided as inputs to the phase-field model in order to simulate location specific microstructures. A wide range of transient nonequilibrium physical phenomena take place within the molten pool (refer to Fig.~\ref{figure}a). Some of these phenomena are fluid flow, Marangoni convection, keyhole mode melting, gravity forces, and recoil pressure due to any evaporation heat losses~\cite{Khairallah2016,King2014,King2015}. A more realistic result can be obtained when all of these complex phenomena are considered. However, this will bring an additional cost, e.g. long computational time and more computational resources. Implementing heat transfer, fluid flow, and Marangoni convection in the melt pool could be a suitable first approach~\cite{Khairallah2016,King2015,Suresh2016}. The resulting mass, momentum, and energy conservation equations can be solved at each discrete element to obtain the transient thermal profiles. For validation of the model, single-track laser scan simulations~\cite{Brandon2017} can be used to generate melt pools with certain dimensions along with the temperature profiles which can be compared with \emph{in situ} thermography measurements~\cite{Trevor2017}. The solidification conditions, for example the cooling rate varies between $10^3$ K s$^{-1}$ and $10^6$ K s$^{-1}$ in different studies due to different approximations of the melt pool physics. An unified modeling approach is therefore needed to quantify the contribution of each melt pool physics and their interactions.
\section{Nanoscale: Molecular dynamics}
Crystal-melt free energy calculations can be performed using suitable interatomic potentials for AM alloys of interest (Ref.~\cite{Hoyt2003} and the references within). Molecular dynamics simulations, via the LAMMPS~\cite{Lammps} software package, are generally conducted to assess the solid-liquid interfacial free energy and kinetic coefficient via a capillary fluctuation technique which monitors the amplitude of atom fluctuations in the interface position~\cite{Hoyt2003} (refer to Fig.~\ref{figure}d). Interface properties data are still rare for the solidification conditions relevant to additive manufacturing regime for multicomponent alloys. Atomistic simulations were also used to estimate the diffusivity of the liquid and the partitioning of the solute across a solid-liquid interface during nonequilibrium rapid directional solidification~\cite{Debierre2000,Asta2011}. MD Simulations can help to estimate the parameters that characterize a solid-liquid interface in AM regime. Recently, MD simulations were performed on a stable aluminum solid-liquid interface under rapid solidification conditions and it was found that the interfacial free energy increased by a factor of 1.25 as the temperature gradient increased by a factor of 3, while the interface anisotropy parameter remained independent of the solidification conditions~\cite{Brown2017}. The interfacial properties of multicomponent alloys under non-equilibrium conditions need to be calculated in MD for use in the phase-field model.
\section{Mesoscale: Phase-field}
Phase-field models use the FEA simulated melt pool solidification conditions and the MD simulated solid-liquid interface information to simulate the solidification microstructures. AM alloys of interest, for example Inconel 718, are typically multicomponent alloys with over a dozen elements. A three component analog, for example Ni$-$Cr$-$Nb, may describe the alloy 718 effectively. It is noteworthy to mention that the design of a ternary AM alloy depends on the microstructural features of interest; two classes of alloys are possible which include either the positive segregation elements (equilibrium partition coefficient $>$ 1) such as Ti, Mo, and Nb in Inconel 718 or the negative segregation elements (equilibrium partition coefficient $<$ 1) such as Fe and Cr. In the former class of alloys, the $\gamma$-dendrite will be lean in Ti, Mo, and Nb and the liquid will be enriched with these elements as the solidification proceeds, while solute partitioning occurs in the opposite direction in the later. A multicomponent phase-field formalism~\cite{steinbach2009,Kundin2015,Abhik2012,Mathis2011,Ghosh2017_eutectic} can be adopted which combines the chemical bulk free energy, the interfacial free energy, and the driving forces for phase transformations such as lowering of the chemical potential. The resultant free energy functional is then minimized using standard variational derivatives with respect to the microstructure field variables, i.e. composition, temperature, and order parameter. The resulting time-dependent equations are known as the Cahn-Hilliard equation~\cite{Cahn1958} which describes the temporal evolution of the conserve quantity composition and the Allen-Cahn equation~\cite{Allen1979} or the time dependent Ginzburg-Landau equation which describes the temporal evolution of the non-conserved quantity order parameter~\cite{boettinger2002,chen2002,steinbach2009}. These partial differential equations are solved in 2D/3D on a uniform/adaptive mesh, using the finite difference/finite volume method, and explicit/semi-implicit time stepping scheme. Simulations often begin with either a thin layer or a circular solid seed in the supercooled liquid at the bottom of the simulation box with relevant initial and boundary conditions and with random, small amplitude perturbations. Stable perturbations grow with time and break into steady state dendritic microstructures (refer to Fig.~\ref{figure}c). For equiaxed mode of solidification, nucleation mechanisms need to be incorporated in the phase-field equations~\cite{Montiel2012}. Software packages such as MATLAB~\cite{Matlab}, PARAVIEW~\cite{Paraview}, GNUPLOT~\cite{Gnuplot} are generally used to visualize and analyze those complex morphologies in the following facets, which need further attention.

\subsection{Dendrite properties}
The size, shape and distribution of dendrites are different at different locations in the melt pool. Phase-field simulations for AM solidification conditions often result in columnar dendrites with primary arms with average spacing $\approx$ \SI{1.0}{\micro\metre} and secondary sidearms with average spacing $\approx$ \SI{0.5}{\micro\metre}~\cite{Murr2012,supriyo2017}. These spacings control the material properties, such as yield strength and tensile strength. A prediction of these spacings from the microstructure simulations and subsequent comparison with the experiments and dendrite growth theories are therefore essential. This comparison can be used, for example, as a standard reference spacing data for Ni$-$based superalloys for AM solidification. Fourier analysis and other methods~\cite{Rappazbook} can be used to extract the dominant spacings in the microstructure. These results will be significant, since current understanding is still largely limited to low-velocity casting solidification regime. Moreover, dendrite spacing varies significantly in the presence of melt pool convection, and thus needs to be studied in AM regime~\cite{Lee2010}.

\subsection{Microstructural anisotropies}
Rapid solidification processing leads to nonequilibrium partitioning of solute atoms in the solid and liquid. As a result, solute gets segregated in the volume of liquid that solidifies in the spaces between the already solidified dendrites and gets enriched by $\approx$ 2 to 5 times of the nominal composition of the alloy element with the progress of solidification~\cite{supriyo2017,Trevor2017}. This is known as microsegregation~\cite{kurzbook}. The presence of multiple elements in the liquid makes the elemental segregation in the interdendritic regions complex. Moreover, the orientation/texture of the growing dendrites is often different in different locations within the melt pool, depending on the direction of the temperature gradient/heat flow. Anisotropy in orientations (the change in preferred orientation with respect to the growth direction), leading to intrinsic anisotropy in the mechanical properties, is therefore natural to consider. An anisotropic gradient energy surface tension~\cite{supriyo2014,ghosh2015} in the phase-field free energy functional and a rotation matrix representation~\cite{supriyo2014,ghosh2015} for the normal vector components of the solid-liquid interface can suitably describe the orientation selection during solidification~\cite{supriyo2014,ghosh2015}. Simulations can also be performed in single/bicrystals in which the misorientation angles and the convergent/divergent growth conditions between columnar dendrites are considered as solidification variables. In this context, the competition and the transition between the columnar and the equiaxed growth modes of dendrites were also modeled using phase-field~\cite{Montiel2012} or stochastic analyses~\cite{Asta2009}, but in the low-velocity limit. A quantitative description of these phenomena in the high-velocity limit will be a significant step towards microstructure and property control, since this knowledge is required to perform the solid-state homogenization heat treatment. A material with uniform properties will therefore be designed and controlled.

\subsection{Solidification defects}
A prediction of the solidification defects such as microporosity, solidification cracking, and residual stresses in the semisolid mushy zone during the late stages of solidification is very important, which have yet to be explored for AM solidification. The mushy zone in columnar dendrites is a two-phase solid and liquid coexistence region between the fully solid and the fully liquid states where majority of the solidification defects form. Those defects arise due to random growth of the solid dendrites toward each other and finally coalesce, leading to insufficient feeding of the liquid to accommodate shrinkage. As the fraction of solid in the mushy zone increases to $\approx$ 0.6 to 0.98~\cite{Rappazbook,Rappaz1999}, the liquid is not able to flow freely and compensate for shrinkage, resulting in microporosity. The semisolid mushy zone therefore becomes weak and ruptures when stressed. This phenomenon is known as hot tearing. This behavior can be quantified by extracting the fraction of solid and liquid from either the simulation data or the Scheil solidification approximation~\cite{Rappaz1999,Rappazbook} and making a correlation with the Euler characteristics~\cite{Hoshenkopelman} of the solidifying sections by extracting the connection topology between the microstructure order parameter, i.e. the coalescence behavior~\cite{supriyo20173d}. One can also determine the mean and Gaussian curvatures at each point of the solid-liquid interface in this purpose~\cite{Voorhees2010,Asta2009,Rappazbook}. Further, the mushy zone can be extracted from the microstructure and can be provided as an input to the volume-of-fluid (VOF) based methods which couple Darcy's law and mass-conservation continuity equations~\cite{Rappaz1999,Wang2004,Rappazbook}. The residual liquid fraction in the mushy zone, primary and secondary dendrite arm spacings, and local solidification conditions can be used for the calculation of criteria functions of linear or low-order polynomial forms for the size, distribution, and growth of the pores. The pore-microstructure interactions can also be modeled in this regard using stochastic approaches for nucleation of pores in combination with continuum solutions for diffusion~\cite{Lee2001}. Once these defect formation mechanisms are analyzed, approaches could be prescribed in order to reduce/eliminate them.

\subsection{Precipitation of solid phases}
Solid-state phase transformations follow the solidification process when precipitation of the secondary solid phases takes place. The simulated microstructures and the representative microstructure variables (composition, temperature, and order parameter) can be used as inputs for the simulation of subsequent solid-state phase changes. As the columnar/equiaxed dendrites grow, as a direct consequence of microsegregation, they leave behind solute enriched pockets in the interdendritic regions (refer to Fig.~\ref{figure}c) in a mechanism similar to the Plateau-Rayleigh instability~\cite{supriyo2017,Trevor2017,Ghosh2018}. Secondary solid phases, such as $\gamma '$, $\gamma ''$, $\delta$, and Laves in Inconel 718~\cite{Knorovsky1989}, are expected to form in these pockets following an eutectic or non-eutectic type of reaction beyond a threshold solute composition and below a certain temperature. On average, the volume fraction of these phases decreases with increasing cooling rate in the melt pool and typically varies between $\approx$ \SI{2}{\%} and \SI{20}{\%} in experiments~\cite{Radhakrishna1997,Zhao2008,Qi2009,Ram2004}. The size and distribution of the secondary phases precipitation in the matrix determine the tensile strength, fracture toughness, and fatigue properties of the solidified material~\cite{Nandwana2017}. A finer size and discrete distribution of those phases are beneficial compared to a coarser and continuous distribution, when resistance to deformation is considered~\cite{Radhakrishna1997,Zhao2008,Qi2009,Ram2004}. Determination of the influence of the solidification conditions on the size, distribution, and volume fraction of the secondary phases can be significantly different in AM solidification compared to casting solidification and is therefore essential. A proper implementation of the threshold composition and temperature boundary conditions in the phase-field model could predict these solid-state phases~\cite{Ghosh2018}. Note that these phases form in the as-deposited microstructures as a function of the solute content or the microsegregation. Following solidification, the as-deposited microstructures often undergo the solution heat treatment, such as annealing, as a function of time and temperature to dissolve the desired/undesired secondary phases completely/partially in the matrix in order to modify the distribution of these phases for improved material properties.

\subsection{Solid state modeling}
Phase-field simulations can also predict the stress evolution during solidification processes by a mechanical coupling of the phase and temperature fields using the stress equilibrium equation with elasto-plastic constitutive equations, the numerical solutions of which reveal the residual stress distribution in the dendritic microstructures~\cite{Khachaturyan,Uehara2008,Steinbach2011}. Such stress distribution can be effectively engineered by the laser processing parameters, which determine the melt pool solidification conditions and ultimately the morphology of the microstructure. Depending on the alloy and the ultimate application of the as-built part, it may be necessary to perform post-solidification thermal processing via homogenization heat treatment and hot isostatic pressing to obtain the desired mechanical properties for optimum performance. Moreover, the formation and growth of the secondary solid precipitates (such as carbides, intermetallics and Laves) need to be modeled~\cite{chen2002,Steinbach2011} to predict the processing window (for example, as a function of time and temperature during heat treatment) of each phases, depending on whether the precipitate is either beneficial or harmful in terms of yield strength, tensile strength, fracture toughness and fatigue life requirements~\cite{Korner2016}. The mechanical interactions, interface anisotropies and orientations, defect densities and diffusion and growth kinetics are significantly different in solids compared to solidification microstructure evolution~\cite{chen2002,Steinbach2011,supriyo_pccp}. The discussion of these topics is beyond the scope of the present article. The present review is limited only to the as-deposited state of the microstructures.

\section{Summary}
Additive manufacturing has the potential to become the technology of the future. There is an increasing demand for the predictive simulation of AM microstructures to achieve better material properties than traditional casting and metal forming routes. Future work is still required to address the process and microstructural challenges during AM to improve the confidence in the quality of the material in service. In this context, the present report reviews the multiscale modeling of multicomponent solidification and solid-state transformations as far as they relate to the solidification models, considering Inconel 718 as an example alloy. The numerical results are also needed to benchmark with experiments and other simulations, since AM microstructural features are far-from-equilibrium. In this way, the processing-microstructure-property correlation of AM materials can be improved which will enable the AM industry to reliably use this novel application to manufacture parts of predictable quality and behavior.
 
\section*{Acknowledgements}
I thank Dilip Banerjee, William Boettinger, Jonathan Guyer, Trevor Keller, Greta Lindwall, Li Ma, Kevin McReynolds, and Nana Ofori-Opoku from National Institute of Standards and Technology, Marianne Francois and Chris Newman from Los Alamos National Laboratory, and Balasubramaniam Radhakrishnan from Oak Ridge National Laboratory for useful discussions.

\section*{References}
\bibliographystyle{unsrt}
\bibliography{papers}
\end{document}